\documentclass{jfm}
\usepackage{color}
\usepackage{amstext}
\usepackage{amssymb}
\usepackage{graphicx}
\usepackage{amsmath}

\newcommand\const{\mathrm{const}}
\newcommand\Div{\mathrm{div}\,}

\newcommand\vX{\boldsymbol{X}}
\newcommand\vP{\boldsymbol{P}}

\newcommand\vV{\boldsymbol{V}}

\newcommand\vf{\boldsymbol{f}}
\newcommand\vh{\boldsymbol{h}}

\newcommand\vu{\boldsymbol{u}}
\newcommand\vv{\boldsymbol{v}}

\newcommand\vx{\boldsymbol{x}}

\newcommand\vp{\boldsymbol{p}}
\newcommand\vq{\boldsymbol{q}}
\newcommand\vQ{\boldsymbol{Q}}
\newcommand\vg{\boldsymbol{g}}

\newcommand\vxi{\boldsymbol{\xi}}

\renewcommand{\theequation}{\arabic{section}.\arabic{equation}}

\begin{document}

 {\title[Hierarchy of Distinguished Limits and Drifts for Oscillating Flows] {Hierarchy of Distinguished Limits and Drifts for Oscillating Flows}}

\author[V. A. Vladimirov]{V.\ns A.\ns V\ls l\ls a\ls d\ls i\ls m\ls i\ls r\ls o\ls v}

\affiliation{DOMAS, Sultan Qaboos University, Oman and DAMTP, University of Cambridge, UK}

\pubyear{2010} \volume{xx} \pagerange{xx-xx}
\date{September 16, 2015 and in revised form ???}

\setcounter{page}{1}\maketitle \thispagestyle{empty}

\begin{abstract}
Lagrangian motions of fluid particles in a general velocity field oscillating in time are studied with the use of the two-timing method.
Our aims are:
(i) to calculate systematically the most general and practically usable asymptotic solutions;
(ii) to check the limits of applicability of the two-timing method by calculating the averaged motion without making any assumptions;
(iii) to classify various drift motions and find their limits of applicability;
(iv) to introduce a logical order into the area under consideration;
(v) to open the gate for application of the same ideas to the studying more complex systems.
Our approach to study a drift is rather unusual: instead of solving the ODE for trajectories we consider a hyperbolic PDE for a scalar lagrangian field $a(\vx,t)$, trajectories represent the characteristics curves for this PDE.
It leads us to purely eulerian description of lagrangian motion, that greatly simplifies the calculations.
There are two small scaling parameters in the problem: a ratio of two time-scales and a dimensionless vibrational amplitude.
It leads us to the sequence of problems for distinguished limits.
We have considered four distinguished limits which differ from each other by the scale of slow time.
We have shown that each distinguished limit produces an infinite number of solutions for $a(\vx,t)$.
A classical drift appears in the case of a purely oscillating flow.
At the same time, we have shown that the concept of drift motion is not sufficient for the description of lagrangian dynamics and some `diffusion'  terms do appear.
Five examples of different options of drifts and `diffusion' are given.
\end{abstract}

\section{Introduction}

It is well-known that in oscillating flows the motion of a material (fluid) particle consists of two parts:
oscillating and non-oscillating. The latter is known as a \emph{drift}, see
\cite{Stokes, Maxwell, Lamb, LH, Darwin, Hunt, Batchelor, Lighthill, McIntyre, Craik, Grimshaw, Craik0,
Benjamin, Hunt1, Eames, Buhler}, and many others. The great significance of a drift due to its key role in the vortex dynamics of oscillating flows and  Langmuir circulations
was recently analyzed from a new perspective by \cite{VladimirovL, VladProc}.
Here we study the advection of a lagrangian scalar field by a given oscillating velocity.
We employ the two-timing approach combined with eulerian time-averaging operation in the form introduced by  \cite{Vladimirov2, Yudovich,
Vladimirov1}. The two-timing method has been used by many authors, see \cite{Nayfeh, Verhulst}, however our analysis goes well beyond the usual calculations of main approximations in various special cases.
Our analytic calculations are straightforward by their nature, but they do include a large number of integration by parts and algebraic transformations;
the performing of such calculations in a general formulation represents a `champion-type' result by itself.
These calculations are too bulky to be presented in JFM, they are described in detail in the arXiv papers by \cite{VladimirovDr1, VladimirovDr2} quoted below as I and II,
therefore here we present only some results of these calculations.
\emph{The main purpose of this paper is to
introduce a new systematic and general viewpoint on the hierarchy of distinguished limits and drifts}.
This new viewpoint helps to organize and unify a number of results, including that of I and II.

\section{Formulation of Problem}

A fluid flow is given by its velocity field $\vv^*(\vx^*,t^*)$, where $\vx^*=(x_1^*,x_2^*,x_3^*)$ and $t^*$ are cartesian coordinates and time, asterisks stand for dimensional variables.
We suppose that this field is sufficiently smooth, but we do not suppose that it satisfies any equations of motion.
The dimensional advection equation for a scalar lagrangian field $a(\vx^*,t^*)$ is
\begin{eqnarray}
a_{t^*}+(\vv^*\cdot\nabla^*) a=0,\qquad a_{t^*}\equiv \partial a/\partial {t^*}
\label{exact-1}
\end{eqnarray}
This equation describes the motions of
a lagrangian marker in either an incompressible or compressible fluid or the advection of a passive scalar admixture
with concentration $a(\vx^*,t^*)$ in an incompressible fluid. In the case of a compressible fluid it also describes the
advection of a passive scalar admixture, where $a$ represents a ratio of concentration of the admixture to
the density of a fluid.
The hyperbolic equation (\ref{exact-1}) has characteristics curves (trajectories) $\vx^*=\vx^*(t^*)$ described by an ODE
\begin{eqnarray}
{d\vx^*}/{dt^*}=\vv^*,\qquad \vx^* |_{t=0}=\vX^*
\label{exact-1a}
\end{eqnarray}
where $\vx^*$ and $\vX^*$ are eulerian and lagrangian coordinates.
The classical description of  drift motion follows after the integration of (\ref{exact-1a}),
however for higher approximations it requires very bulky operations with lagrangian displacements.
Therefore we solve the equation (\ref{exact-1}) that allows to use the eulerian average operation
and to simplify calculations. The oscillating field $\vv^*$ possesses the characteristic scales of velocity $U$, length $L$ and frequency $\omega^*$.
These three parameters give a Strouhal number $St= L\omega^*/U$, hence the dimensionless variables and parameters (written without asterisks) are not unique; we use the following set
\begin{eqnarray}
&& \vx^*= L\vx,\ t^*=(L/U)t,\ \omega^*=(U/L)\omega,\ \vv^*=U\omega^\beta\vu=(St)^\beta U\vu;\ \beta=\const<1\label{variables}
\end{eqnarray}
We accept that $\vu$ has the `two-timing' functional form
\begin{eqnarray}
&& \vu(\vx, s, \tau),\qquad \tau=\omega t,\quad s= t/\omega^\alpha;\quad \vu\sim O(1), \quad \alpha=\const >-1 \label{exact-2}
\end{eqnarray}
where $\tau$-dependence is always $2\pi$-periodic, while $s$-dependence is arbitrary.
Two indefinite constants  $\alpha$ and $\beta$ will be defined in distinguished limits.
The restriction $\alpha>-1$  makes the variable $s$ `slow' in comparison with $\tau$,  while $\beta<1$ gives a small vibrational spatial amplitude; $\omega\equiv St$ is considered as a large parameter.

In dimensionless  variables (after the use of chain rule) (\ref{exact-1}) takes the form
\begin{eqnarray}
&&\omega a_\tau+ \frac{1}{\omega^\alpha} a_s+\omega^\beta({\vu}\cdot\nabla)a= 0, \qquad {\partial}/{\partial{t}}=\omega\partial/\partial\tau+\frac{1}{\omega^\alpha}\partial/\partial s\label{exact-6}
\end{eqnarray}
where the subscripts $\tau$ and $s$  stand for partial derivatives; $s$ and $\tau$ represent two mutually dependent time-variables, which are called \emph{slow time} and \emph{fast time}.
The functional structure of $\vu$ (\ref{exact-2}) underpins the key suggestion that the solution of (\ref{exact-1})
 has the same structure
\begin{eqnarray}
&& a=a(\vx, s, \tau)\label{exact-3}
\end{eqnarray}
Equation (\ref{exact-6}) can be rewritten in the form containing two independent small parameters
\begin{eqnarray}
&&a_\tau+\varepsilon_2 a_{s}+\varepsilon_1 \vu\cdot\nabla a= 0;\quad \varepsilon_1\equiv \omega^{(\beta-1)}, \quad \varepsilon_2\equiv 1/\omega^{(\alpha+1)}
\label{exact-6a}
\end{eqnarray}
Hence, we operate in the plane $(\varepsilon_1, \varepsilon_1)$, where we study asymptotic limit $(\varepsilon_1,\varepsilon_2)\to (0,0)$.
Different asymptotic paths $(\varepsilon_1,\varepsilon_2)\to (0,0)$ may produce different solutions, such paths are called distinguished limits.
In \eqref{exact-6a},    following the methodology of two-timing approach, we (temporary) consider $\tau$ and $s$ as independent variables.

In order to make further analytic progress we require some specific notations.
In this paper we assume that any dimensionless function $f(\vx,s,\tau)$ has the following properties:
(i)  $f\sim {O}(1)$ and  all its required $\vx$-, $s$-, and $\tau$-derivatives are also ${O}(1)$;
(ii) is $2\pi$-periodic in $\tau$, i.e.\ $f(\vx, s, \tau)=f(\vx,s,\tau+2\pi)$;
(iii) has an average given by
$$
\langle {f}\,\rangle \equiv \frac{1}{2\pi}\int_{\tau_0}^{\tau_0+2\pi}
f(\vx, s, \tau)\, d \tau\equiv \overline{f}(\vx,s) \qquad \forall\ \tau_0=\const;
$$
(iii) can be split into averaged and purely oscillating parts, $f(\vx, s, \tau)=\overline{f}(\vx, s)+\widetilde{f}(\vx, s, \tau)$ where the \emph{tilde-functions} (or purely oscillating functions) are such that $\langle \widetilde f\, \rangle =0$ and the \emph{bar-functions} $\overline{f}(\vx,s)$ are $\tau$-independent;
(iv)  we introduce a special notation for  the
\emph{tilde-integration}, which keeps the result of integration in the tilde-class:
\begin{equation}
\widetilde{f}^{\tau}\equiv\int_0^\tau \widetilde{f}(\vx,s,\tau')\, d \tau'
-\frac{1}{2\pi}\int_0^{2\pi}\Bigl(\int_0^\mu
\widetilde{f}(\vx,s,\tau')\, d \tau'\Bigr)\, d \mu.
\label{ti-integr}
\end{equation}

\section{Distinguished Limits and Related Solutions}
It can be shown (see I,II) that there are series of distinguished limits for the equation (\ref{exact-6a}), which represent the one-parametric distinguished path $\varepsilon_1=\varepsilon$, $\varepsilon_2=\varepsilon^n$
with an integer $n>0$:
\begin{eqnarray}
&&a_\tau+\varepsilon^n a_{s}+\varepsilon \vu\cdot\nabla a= 0
\label{exact-6c}
\end{eqnarray}
Let us consider first four cases of Distinguished Limits DL($n$) with $n=1,2,3,4$:

DL(1)  $s= t$;  $\vu=\overline{\vu}(\vx,s)+\widetilde{\vu}(\vx,s,\tau)$;

DL(2) $s=\varepsilon t$;\  $\vu=\widetilde{\vu}(\vx,s,\tau)$;

DL(3)  $s=\varepsilon^2 t$;\  $\vu=\widetilde{\vu}(\vx,s,\tau)$ which satisfies condition $\overline{\vV}_0\equiv 0$;

DL(4) $s=\varepsilon^3 t$;\  $\vu=\widetilde{\vu}(\vx,s,\tau)$ which satisfies conditions $\overline{\vV}_0\equiv 0$, $\overline{\vV}_1\equiv 0$.

The used notations are
\begin{eqnarray}
&&\vV_0\equiv \frac{1}{2}[\widetilde{\vu},\widetilde{\vxi}],\quad
\overline{\vV}_1\equiv\frac{1}{3}\langle[[\widetilde{\vu},\widetilde{\vxi}],\widetilde{\vxi}]\rangle,\quad \widetilde{\vxi}\equiv\widetilde{\vu}^\tau
\label{4.18}
\end{eqnarray}
where the commutator $[\vf,\vg]\equiv(\vg\cdot\nabla)\vf-(\vf\cdot\nabla)\vg$ for any vector-functions $\vf$ and $\vg$. In all cases we are looking for the solution in the form of regular series
\begin{eqnarray}
a(\vx,t,\tau)=\sum_{k=0}^\infty\varepsilon^k a_k(\vx,t,\tau),\quad k=0,1,2,\dots
\label{basic-4aa}
\end{eqnarray}

The case DL(1) with $\overline{\vu}=O(1)\neq 0$  naturally corresponds to the advection speed of order one, hence $s=t$ and the averaged equation of zero approximation is
\begin{eqnarray}\nonumber
\left(\partial_s+ \overline{\vu}\cdot\nabla\right)\overline{a}_{0}=0.
\end{eqnarray}
Here we do not consider here this case in detail, one can find it in I.

The most instructive from our point of view is the case of DL(2).
%In DL(1) equations of higher approximations all the coefficients are similar to that of DL(2) (which are given below), just they appear in the equations which are one order in $\varepsilon$ higher than in DL(2).
Here we have a longer slow time  scale $s=\varepsilon t$, since here $\overline{\vu}\equiv 0$ and, naturally, the speed of
advection is lower.
The substitution of (\ref{basic-4aa}) into (\ref{exact-6c}) and a number of transformations lead to the averaged equations of successive approximations, which are derived in
I, II
\begin{eqnarray}
&&\left(\partial_s+ \overline{\vV}_0\cdot\nabla\right)\overline{a}_{0}=0,\quad \partial_s\equiv \partial/\partial s\label{4.15}\\
&&\left(\partial_s+ \overline{\vV}_0\cdot\nabla\right)
\overline{a}_{1}+(\overline{\vV}_1\cdot\nabla)\overline{a}_0=0\label{4.16}\\
&&\left(\partial_s+ \overline{\vV}_0\cdot\nabla\right)\overline{a}_{2}+
(\overline{\vV}_1\cdot\nabla)\overline{a}_1+(\overline{\vV}_2\cdot\nabla)\overline{a}_0=
\frac{\partial}{\partial x_i}\left(
\overline{\chi}_{ik}\frac{\partial\overline{a}_0}{\partial x_k}\right),
\label{4.17}\\
&&\overline{\vV}_2\equiv\frac{1}{4}\langle[[\widehat{\vV}_0,\widetilde{\vxi}],\widetilde{\vxi}]\rangle +
\frac{1}{2}\langle[\widetilde{\vV}_0,\widetilde{\vV}_0^\tau]\rangle+
\frac{1}{2}\langle[\widetilde{\vxi},\widetilde{\vxi}_t]\rangle
+\frac{1}{2}\langle\widetilde{\vxi}\Div\widetilde{\vu}'+ \widetilde{\vu}'\Div\widetilde{\vxi}\rangle,\label{4.19}
\\
&&\widetilde{\vu}'\equiv\widetilde{\vxi}_t-[\overline{\vV}_0,\widetilde{\vxi}],
\label{4.20a}\\
&&2\overline{\chi}_{ik}\equiv\langle\widetilde{u'}_i\widetilde{\xi}_k+\widetilde{u'}_k\widetilde{\xi}_i\rangle=
\mathfrak{L}_{\overline{\vV}_0}\langle\widetilde{\xi}_i\widetilde{\xi}_k\rangle,\label{4.20}\\
&&\mathfrak{L}_{\overline{\vV}_0}\overline{f}_{ik}\equiv
\left(\partial_s+
\overline{\vV}_0\cdot\nabla\right)\overline{f}_{ik}-\frac{\partial\overline{V}_{0k}}{\partial x_m}\overline{f}_{im}-
\frac{\partial\overline{V}_{0i}}{\partial x_m}\overline{f}_{km}
\label{4.20b}
\end{eqnarray}
where the operator $\mathfrak{L}_{\overline{\vV}_0}$ is such that $\mathfrak{L}_{\overline{\vV}_0}
\overline{f}_{ik}=0$ represents the condition for tensorial field $\overline{f}_{ik}(\vx,t)$
to be `frozen' into $\overline{\vV}_0(\vx,t)$ (also known as the Lie derivative). Three equations (\ref{4.15})-(\ref{4.17}) can be written
as a single advection-pseudodiffusion equation (with an error ${O}(\varepsilon^3)$)
\begin{eqnarray}
&&\left(\partial_s+ \overline{\vV}\cdot\nabla\right)\overline{a} =
\frac{\partial}{\partial x_i}\left(\overline{\kappa}_{ik}\frac{\partial\overline{a}}{\partial x_k}\right)
\label{4.21}\\
&&\overline{\vV}=\overline{\vV}^{[2]}=\overline{\vV}_0+\varepsilon\overline{\vV}_1+\varepsilon^2
\overline{\vV}_2,\label{4.22}\\
&&\overline{\kappa}_{ik}=\overline{\chi}_{ik}^{[2]}=\varepsilon^2\overline{\chi}_{ik}\label{4.23}\\
&&\overline{a}=\overline{a}^{[2]}=\overline{a}_0+\varepsilon\overline{a}_1+\varepsilon^2\overline{a}_2
\label{4.22aa}
\end{eqnarray}
Eqn. (\ref{4.21}) shows that the averaged motion represents a drift with velocity $\overline{\vV}=\overline{\vV}^{[2]}+{O}(\varepsilon^3)$ and
\emph{pseudodiffusion} with matrix coefficients
$\overline{\kappa}_{ik}=\varepsilon^2\overline{\chi}_{ik}+{O}(\varepsilon^3)$.
We have introduced the term  {\emph{pseudodiffusion}} on the ground of the following:
(i) the evolution of $\overline{a}$ is described by an advection-diffusion-type equation (\ref{4.21}); (ii) the
diffusion type term represents a
known source-type term in the second approximation; and (iii) the equation (\ref{4.21}) is valid only for regular
asymptotic expansions (\ref{4.22})-(\ref{4.22aa}).

For the case DL(3) we impose a restriction $\overline{\vV}_0\equiv 0$ and derive equations (see I):
\begin{eqnarray}
&&\left(\partial_s+ \overline{\vV}_1\cdot\nabla\right)\overline{a}_{0}=0\label{4.15a}\\
&&\left(\partial_s+ \overline{\vV}_1\cdot\nabla\right)\overline{a}_{1}+
(\overline{\vV}_2\cdot\nabla)\overline{a}_0=
\frac{\partial}{\partial x_i}\left(
\overline{\chi}_{ik}\frac{\partial\overline{a}_0}{\partial x_k}\right),
\label{4.17a}\\
&&\overline{\vV}_2\equiv\frac{1}{4}\langle[[\widetilde{{\vV}}_0,\widetilde{\vxi}],\widetilde{\vxi}]\rangle +
\frac{1}{2}\langle[\widetilde{\vV}_0,\widetilde{\vV}_0^\tau]\rangle+
\frac{1}{2}\langle[\widetilde{\vxi},\widetilde{\vxi}_s]\rangle
+\frac{1}{2}\partial_s\langle\widetilde{\vxi}\Div\widetilde{\vxi}\rangle,\label{4.19a}\\
&&2\overline{\chi}_{ik}=
\partial_s\langle\widetilde{\xi}_i\widetilde{\xi}_k\rangle\label{4.20b}
\end{eqnarray}

For the case DL(4) we impose two restrictions  $\overline{\vV}_0\equiv 0$ and $\overline{\vV}_1\equiv 0$ and derive the equation (see I):
\begin{eqnarray}
&&\left(\partial_s+ \overline{\vV}_2\cdot\nabla\right)\overline{a}_{0}=
\frac{\partial}{\partial x_i}\left(
\overline{\chi}_{ik}\frac{\partial\overline{a}_0}{\partial x_k}\right),
\label{4.17c}
\end{eqnarray}
with the same $\overline{\vV}_2$ and $\overline{\chi}_{ik}$ as in DL3.
The comparison between the averaged equations for DL(1)--DL(4) shows that the same coefficients from higher approximations in DL(N)  appear in DL(N-1) in the equations of the previous order in $\varepsilon$.
The higher approximations DL(5), \emph{etc.}) can be derived similarly, however the calculations become extremely cumbersome.

One should recognize, that in all presented distinguished limit solutions the meaning of key parameter $\varepsilon$ has not been fully defined yet.
To make a progress in this direction, let us again consider DL(2) for (\ref{exact-6c})
\begin{eqnarray}
&&a_\tau+\varepsilon^2 a_{s}+\varepsilon \widetilde{\vu}\cdot\nabla a= 0
\label{exact-6d}
\end{eqnarray}
In order to make this equation coinciding with (\ref{exact-6}) we must choose $\alpha$ and $\beta$ as:
\begin{equation}
\beta=(1-\alpha)/2,\quad \varepsilon=1/\omega^{(\alpha+1)/2}.
\label{main-eq111}
\end{equation}
Now, for any solution of (\ref{exact-6d}), (which represents a distinguished limit solution) we obtain an infinite number of solutions, one solution for any real number $\alpha>-1$.
Those solutions correspond to different slow time variables and different magnitudes of given velocity. For example:

 $\bullet$ If $\widetilde{\vu}$ is given as function of variables $\tau=\omega t$ and $s=t$, then we must take $\alpha=0$, $\beta=1/2$, hence a dimensionless velocity $(St)^\beta \widetilde{\vu}$ in (\ref{variables}) is of order $O(\sqrt{\varepsilon})$, and $\varepsilon=1/\sqrt\omega$.

$\bullet$ The most frequently considered case, when a velocity is of order one, corresponds to $\alpha=1$, $\beta=0$, and $\varepsilon=1/\omega$.

$\bullet$ Another interesting possibility corresponds to the case $\alpha=\beta =1/3$. Here, $s=t/\sqrt[3]\varepsilon$, $\varepsilon=\omega^{-2/3}$ and velocity $(St)^\beta \widetilde{\vu}=\sqrt[3]\omega\widetilde{\vu}$. Although such a scaling may look exotic, it is required if a particular slow time-scale $s=t/\sqrt[3]\omega$ is prescribed in $\widetilde{\vu}$.

$\bullet$ Our results for DL(2) are particularly striking for the case of $\widetilde{\vu}(\vx,\tau)$ not depending on $s$, which represents a flow with purely periodic velocity oscillations. As one can see from the above consideration, the scale of slow time in this case is uniquely determined by given velocity $\omega^{\beta}\widetilde{\vu}$. For every value of $\beta$ we have the related slow time $s=t/\omega^\alpha$. It must be accepted, that solutions with different $\beta$ are physically different.

$\bullet$ The general tendency given by \eqref{main-eq111} is physically natural: in order to shorten the slow time-scale (decreasing $\alpha$), one needs to increase the amplitude of a given velocity (increasing $\beta$).

$\bullet$ In general, transformations similar to (\ref{exact-6d}),(\ref{main-eq111}) produce an infinite number of different solutions for each case DL(1-4).

\section{Examples}

%Let us now consider several examples of velocity fields.

\emph{Example 1.} \emph{The superposition of two modulated oscillatory fields of the same frequency}.
\begin{eqnarray}
&&\widetilde{\vu}(\vx,s,\tau)=\overline{\vp}(\vx,s)\sin\tau+
\overline{\vq}(\vx,s)\cos\tau  \label{Example-1-1}
\end{eqnarray}
where
$\overline{\vp}$ and $\overline{\vq}$ are arbitrary bar-functions. The straightforward calculations yield
$
[\widetilde{\vu},\widetilde{\vxi}]=[\overline{\vp},\overline{\vq}].
$
 The drift velocities (\ref{4.18}), (\ref{4.19}) are
\begin{eqnarray}
&&\overline{\vV}_0=\frac{1}{2}[\overline{\vp},\overline{\vq}],\quad
\overline{\vV}_1\equiv 0,
\label{Example-1-4}\\
&&\overline{\vV}_2=\frac{1}{8}\left([\overline{\vP},\overline{\vp}]+[\overline{\vQ},\overline{\vq}]\right)
-\frac{1}{4}\left([\overline{\vp}_s,\overline{\vp}]+[\overline{\vq}_s,\overline{\vq}]\right)+\label{Example-1-4a}\\
&&+\frac{1}{4}\left(\overline{\vp}\,\Div\overline{\vP}'+\overline{\vq}\Div\overline{\vQ}'
+\overline{\vP}'\Div\overline{\vp}+\overline{\vQ}'\Div\overline{\vq}\right),\nonumber\\
&&\overline{\vP}\equiv[\overline{\vV}_0,\overline{\vp}],\quad
\overline{\vQ}\equiv[\overline{\vV}_0,\overline{\vq}],\quad
\overline{\vP}'\equiv\overline{\vp}_s-\overline{\vP},\quad
\overline{\vQ}'\equiv\overline{\vq}_s-\overline{\vQ},\nonumber\\
%&&\widetilde{\vu}'=-\overline{\vP}'\cos\tau+\overline{\vQ}'\sin\tau,\quad\nonumber\\
&&\langle\widetilde{\xi}_i\widetilde{\xi}_k\rangle=
\frac{1}{2}(\overline{p}_i \overline{p}_k + \overline{q}_i \overline{q}_k)\label{Example-1-5}
\end{eqnarray}
Pseudo-diffusion matrix $\overline{\kappa}_{ik}$ follows after the substitution of (\ref{Example-1-5}) into
(\ref{4.20}). The velocity $\widetilde{\vu}$ (\ref{Example-1-1}) is general enough to produce any given function
$\overline{\vV}_0(\vx,t)$. To obtain $\overline{\vp}(\vx,t)$ and $\overline{\vq}(\vx,t)$ one has to solve the equation
\begin{eqnarray}\label{bi-linear}
[\overline{\vp},\overline{\vq}]=(\overline{\vq}\cdot\nabla)\overline{\vp}-
(\overline{\vp}\cdot\nabla)\overline{\vq}=2\overline{\vV}_0(\vx,s)
\end{eqnarray}
which represents an undetermined bi-linear PDE-problem for two unknown functions.

\emph{Example 2.} \emph{ Stokes drift}.

The dimensionless plane velocity field is (see \cite{Stokes, Lamb, Debnath})
\begin{eqnarray}
\widetilde{\vu}=Ae^{ky}\left(\begin{array} {c} \cos(kx-\tau)\\ \sin (kx-\tau)\end{array}\right),\quad (x,y)\equiv(x_1,x_2)
\label{Example-3-1}
\end{eqnarray}
where one can choose $A=1$ and $k=1$; however, let us keep both $A$ and $k$ in the formulae in order to trace the
physical meaning. The fields $\overline{\vp}(x,y)$, $\overline{\vq}(x,y)$ (\ref{Example-1-1}) are
\begin{eqnarray}
\overline{\vp}=Ae^{ky}\left(\begin{array}{c} \sin kx\\ -\cos kx\end{array}\right),\quad
\overline{\vq}=Ae^{ky}\left(\begin{array}{c} \cos kx\\ \sin kx\end{array}\right)
\label{Example-3-2}
\end{eqnarray}
The calculations of (\ref{Example-1-4}) yield
\begin{eqnarray}
\overline{{\vV}}_0=k A^2 e^{2ky}\left(\begin{array}{c} 1\\ 0\end{array}\right), \quad  \overline{{\vV}}_1\equiv 0
\label{Example-3-3}
\end{eqnarray}
which represents the classical Stokes drift and a zero first correction
to it.  For brevity, the explicit formula for $\overline{{\vV}}_2$  is not given here.
Further calculations show that
\begin{eqnarray}\nonumber
&&\overline{\chi}_{ik}=-\overline{\chi}
\begin{pmatrix}
0 & 1\\
1 & 0
\end{pmatrix},\quad\text{with}\quad \overline{\chi}\equiv\frac{1}{4}k^2 A^4 e^{3ky}
\end{eqnarray}
One can see that the eigenvalues $\overline{\chi}_1=-\overline{\chi}$ and $\overline{\chi}_2=\overline{\chi}$
correspond to a strongly anisotropic `diffusion'.
The averaged equation (\ref{4.21}) (with an error $O(\varepsilon^3)$) can be written as
\begin{eqnarray}
&&\overline{a}_{t}+(\overline{V}_0+\varepsilon^2 \overline{V}_2) \overline{a}_x=
    \varepsilon^2(\overline{\chi}_y \overline{a}_{x}+\overline{\chi}\, \overline{a}_{xy})\label{Example-3-5}\\
    &&\overline{a}=\overline{a}_0+\varepsilon\overline{a}_1+\varepsilon^2\overline{a}_2\nonumber
\end{eqnarray}
where $\overline{V}_0$ and $\overline{V}_2$ are the $x$-components of corresponding velocities (their $y$-components
vanish). This equation has an exact solution $\overline{a}=\overline{a}(y)$ where $\overline{a}(y)$ id an arbitrary function, which is not effected  by pseudodiffusion.

\emph{Example 3.}  \emph{A spherical `acoustic' wave.}

A velocity potential for an outgoing spherical wave is
\begin{eqnarray}
&& \widetilde{\phi}=\frac{A}{r}\sin(kr-\tau)\label{Example-4-1}
\end{eqnarray}
where $A$, $k$, and $r$ are amplitude, wavenumber, and radius in a spherical coordinate system. The velocity is
purely radial and has a form (\ref{Example-1-1})
\begin{eqnarray}
&& \widetilde{u}=\overline{p}\sin\tau+\overline{q}\cos\tau,\\
&&\overline{p}=A\left(\frac{1}{r^2}\cos kr+\frac{k}{r}\sin kr\right),\quad\overline{q}=A\left(-\frac{1}{r^2}\sin
kr+\frac{k}{r}\cos kr\right)\label{Example-4-1a}
\end{eqnarray}
where $\widetilde{u}, \overline{p}$, and $\overline{q}$ are radial components of corresponding vector-fields. The
fields $\widetilde{\vxi}$ and $[\widetilde{\vu},\widetilde{\vxi}]$ are also purely radial; the radial component for the
commutator is
\begin{eqnarray}
&& \widetilde{\xi} \widetilde{u}_r-\widetilde{u}\widetilde{\xi}_r=A^2 k^3/r^2\label{Example-4-3}
\end{eqnarray}
where $\widetilde{\xi}$ is radial component of $\widetilde{\vxi}$ and subscript $r$ stands for the radial derivative. The drift
(\ref{Example-1-4}) is purely radial with
\begin{eqnarray}
&&\overline{V}_0=\frac{A^2 k^3}{2\,r^2},\quad
\overline{V}_1=0,\quad\overline{V}_2=\frac{A^4k^5}{16r^4}\left(3k^2-\frac{5}{r^2}\right)
\label{Example-4-5}
\end{eqnarray}
It is interesting that $\overline{V}_0$ formally coincides with the velocity, caused by a point source in an
incompressible fluid, and for small $r$ the value of  $\overline{V}_2$
dominates over $\overline{V}_0$, so the series is likely to be diverging. Further calculations yield
\begin{eqnarray}
&&\langle\xi^2\rangle=\frac{A^2}{2r^2}(k^2+1/r^2),\quad \overline{\chi}=A^4k^5/4r^2>0
\label{Example-4-6}
\end{eqnarray}
where $\overline{\chi}$ stands for the only nonzero $rr$-component of $\overline{\chi}_{ik}$. One can see that in this
case pseudodiffusion appears as ordinary diffusion.

\emph{Example 4.}  \emph{ $\overline{\vV}_1$-drift.}

It is interesting to see such flows for which $\overline{\vV}_0\equiv 0$ but $\overline{\vV}_1\neq 0$. Let a velocity field
be a superposition of two standing waves of frequencies $\omega$ and $2\omega$:
\begin{eqnarray}
&&\widetilde{\vu}(\vx,t,\tau)=\overline{\vp}(\vx,t)\sin\tau+\overline{\vq}(\vx,t)\sin 2\tau  \label{Example-5-1}\\
&&[\widetilde{\vu},\widetilde{\vxi}]=\frac{1}{2}[\overline{\vp},\overline{\vq}](2\cos\tau\sin 2\tau-\cos
2\tau\sin\tau)\label{Example-5-3}
\end{eqnarray}
Hence (\ref{4.18}) yields
\begin{eqnarray}\label{Example-5-4}
\overline{\vV}_0=\frac{1}{2}\langle[\widetilde{\vu},\widetilde{\vxi}]\rangle\equiv 0,\quad
\overline{\vV}_1=\frac{1}{3}\langle[[\widetilde{\vu},\widetilde{\vxi}],\widetilde{\vxi}]\rangle=
\frac{1}{8}[[\overline{\vp},\overline{\vq}],\overline{\vp}]
\end{eqnarray}
These expressions produce infinitely many examples of flows with $\overline{\vV}_1$-drift.

\emph{Example 5}  \emph{ Chaotic dynamics for $\overline{\vV}_0$.}

Let a solenoidal/incompressible velocity (\ref{Example-1-1})
be
\begin{eqnarray}
\overline{\vp}=\left(\begin{array}{c}
  \cos y \\
  0 \\
  \sin y
\end{array}\right), \qquad
\overline{\vq}=\left(\begin{array}{c}
  a\sin z \\
  b\sin x+a\cos z \\
  b\cos x
\end{array}\right)
\end{eqnarray}
where  $(x,y,z)$ are  Cartesian coordinates; $a, b$ are constants. Either of this fields, taken separately, produces
simple integrable dynamics of particles. Calculations yield
\begin{eqnarray}
\overline{\vV}_0=\left(\begin{array}{c}
  -a\sin y\sin x-2b\sin y\cos z \\
  b\sin z\sin y-a\cos x\cos y \\
  b\cos z\cos y+2a\sin x\cos y
\end{array}\right),
\end{eqnarray}
Computations of solutions for $d\overline{\vx}_0/ds=\overline{\vV}_0$ which is following from (\ref{exact-1a}) for DL(2) (see I) were performed by Prof. A.B.Morgulis (private communications). He has shown that this steady averaged flow exhibits chaotic dynamics of particles.
In particular, positive
Lyapunov exponents have been observed. Hence, the drift created by a simple oscillatory field can
produce complex averaged lagrangian dynamics.

\section{Discussion}

1. Our study is aimed to create a general viewpoint on distinguished limits and drifts based on the two-timing method. We do not use any additional suggestion and assumptions, hence any physical failure of our results can be related to the insufficiency of two time-scales only. Indeed, the presence of scaling parameters, such as $St$, $\varepsilon_1$, and $\varepsilon_2$, allows to introduce an infinite number of additional time-scales.
From this perspective the problem we study can be seen as a test for sufficiency of the two-timing method.
In particular, some secular (in $s$) terms could appear in solutions of \eqref{4.17} and \eqref{4.17a}.
If it is recognised as unacceptable, then further time-scales should be introduced, which require the systematic development of three-timing \emph{etc.} methods.
Alternatively, one can suggest that the two-timing method fails at the orders of approximations, where the pseudo-diffusion appears.

2. One can rewrite the approximate solution \eqref{4.22aa} back to variable $t$ (along with its tilde-part, see I and II) and substitute it back into the exact original equation (\ref{exact-1}); then a small residual (a nonzero right-hand-side) appears.
The method used allows to produce an approximate solution with a residual
as small as needed. However, the next logical step is more challenging: one has to prove that a solution with a
small residual is close to the exact one.
Such proofs had been performed by \cite{Simonenko, Levenshtam} for the case of vibrational convection. Similar justifications
for other oscillatory flows are not available yet.

3. All results of this paper have been obtained for the class of $\tau$-periodic functions, which is self-consistent. One can consider more general
classes of quasi-periodic, non-periodic, or chaotic solutions. The discussion on this topic is given in I, II.
It is worth to understand the properties of periodic oscillations, in order not to link these properties exclusively to more general solutions.
From other side, the simplicity of calculations for the  $\tau$-periodic solutions allows to obtain very advanced results that can be searched and generalized for more general solutions.

4. The appearance of pseudo-diffusion in the DL(1-4) is especially interesting to study.
Its physical mechanism is discussed in detail in I, II.
It is worth to mention that the averaged pseudo-diffusion coefficient for different flows can be positive, negative, or it can change sign in space and time.

5. \emph{Example 5} demonstrates that a drift can produce chaotic averaged dynamics of particles.
This result leads to numerous new questions and opportunities such as: (i) what is
the relationship between chaotic motions for an original dynamical system and for the averaged one?  (ii) how can a chaotic drift and  pseudo-diffusion complement each other?
(iii) how a chaotic drift can be used in the theory of mixing? (iv) since the averaged dynamics is chaotic, then the related results by
\cite{Arnold,Aref,Ottino,Wiggins,Chierchia} can be efficiently used.

6. It is worth to calculate the characteristics/tragectories directly by solving (\ref{exact-1a}) with the use of the same two-timing method and to compare the related solutions
for drifts with the results presented above. Such calculations are partially presented in I, II.

7. Similar results for a passive vectorial admixture are presented in I, II.
They are closely linked to the problem of kinematic $MHD$-dynamo, see \cite{Moffatt}.
It is physically apparent, that for the majority of shear drift velocities $\overline{\vV}(\vx,t)$ the stretching of material
elements  produces the linear growth of a magnetic field $|\overline{\vh}|\sim t$. At the same time, for the averaged flows with
exponential stretching of `averaged' material lines these examples will inevitably show the exponential growth.

\begin{acknowledgments}
The author is
grateful to Profs. A.D.D.Craik, R.Grimshaw, K.I.Ilin, M.E.McIntyre, H.K.Moffatt, A.B.Morgulis, T.J.Pedley, M.R.E.Proctor, and D.W.Hughes  for helpful discussions.
\end{acknowledgments}

\end{document}